# Emerging chiral optics from chiral interfaces

Xinyan Zhang[1,2,*], Yuhan Zhong[1,2,*], Tony Low[3,†], Hongsheng Chen[1,2,‡], and Xiao Lin[1,2,§]

[1]*Interdisciplinary Center for Quantum Information, State Key Laboratory of Modern Optical Instrumentation, ZJU-Hangzhou Global Science and Technology Innovation Center, College of Information Science and Electronic Engineering, Zhejiang University, Hangzhou 310027, China.*
[2]*International Joint Innovation Center, ZJU-UIUC Institute, Zhejiang University, Haining 314400, China.*
[3]*Department of Electrical and Computer Engineering, University of Minnesota, Minneapolis, Minnesota 55455, USA.*

**Twisted atomic bilayers are emerging platforms for manipulating chiral light-matter interaction at the extreme nanoscale, due to their inherent magnetoelectric responses induced by the finite twist angle and quantum interlayer coupling between the atomic layers. Recent studies have reported the direct correspondence between twisted atomic bilayers and chiral metasurfaces, which features a chiral surface conductivity, in addition to the electric and magnetic surface conductivities. However, far-field chiral optics in light of these consitututive conductivities remains unexplored. Within the framework of the full Maxwell equations, we find that the chiral surface conductivity can be exploited to realize perfect polarization transformation between linearly polarized light. Remarkably, such an exotic chiral phenomenon can occur either for the reflected or transmitted light.**

Due to the pioneering discovery of superconducting and correlated insulating states in twisted bilayer graphene [1-5], twisted atomic bilayers have ignited enormous research interest both in practical applications and fundamental physics [6-11], including optics with 2D twistronics [12-22]. Due to the rotational misalignment and the strong quantum coupling between the neighboring atomic layers, all mirror symmetry in twisted atomic bilayers is broken, and the corresponding light-matter interaction is inherently chiral. As such, twisted atomic bilayers are emerging atomically-thin platforms for chiral optics [23-29] and chiral plasmonics [30-36], which may facilitate novel on-chip applications such as the discrimination of chiral molecules with opposite handedness.

On the other hand, recent studies [35] have revealed the direct correspondence between twisted atomic bilayers and chiral metasurfaces in the framework of the full Maxwell equations. In essence, the twisted atomic bilayer is equivalent to a chiral metasurface, which simultaneously possesses the electric surface conductivity $\bar{\bar{\sigma}}_\mathrm{e}$, magnetic surface conductivity $\bar{\bar{\sigma}}_\mathrm{m}$, and chiral surface conductivity $\bar{\bar{\sigma}}_\chi$. Mathematically, the electromagnetic boundary conditions for such a chiral metasurface are described by [35]:

$$\hat{n} \times (\bar{E}_1 - \bar{E}_2) = -\bar{\bar{\sigma}}_\mathrm{m}(\bar{H}_1 + \bar{H}_2) + \bar{\bar{\sigma}}_\chi(\bar{E}_1 + \bar{E}_2), \qquad (1)$$

$$\hat{n} \times (\bar{H}_1 - \bar{H}_2) = +\bar{\bar{\sigma}}_\mathrm{e}(\bar{E}_1 + \bar{E}_2) + \bar{\bar{\sigma}}_\chi(\bar{H}_1 + \bar{H}_2), \qquad (2)$$

where $\bar{E}_1$ and $\bar{E}_2$ ($\bar{H}_1$ and $\bar{H}_2$) are the components of electric (magnetic) fields parallel to the boundary in the regions above the boundary (i.e., denoted as region 1) and beneath the boundary (region 2), respectively, and $\hat{n} = -\hat{z}$ is the surface normal [Fig. 1]. In principle, these surface conductivities can be arbitrary tensors, which can be constructed by stacking appropriate atomic layers (e.g. in-plane anisotropic or magnetic 2D materials) [7-9,37-39], or with regular metasurface approaches (e.g. nano-patterning) [40-43]. Recent study of chiral plasmonics as governed by equations (1, 2) [35] revealed that the chiral surface conductivity can be exploited to manifest the longitudinal spin of surface plasmons, in addition to the conventional transverse spin of surface plasmons [44-47]. Apart from this work, the field of chiral optics with twisted atomic bilayers remains relatively elusive, and exotic chiral optical phenomena remain to be discovered.

Here, we systematically investigate the far-field chiral optical phenomena with twisted atomic bilayers due to the interplay between the chiral, electric and magnetic surface conductivities. The general expressions for the reflection and transmission coefficients under the incidence of linearly polarized light are analytically derived. Without loss of generality and in accordance with Ref. [35], we set $\bar{\bar{\sigma}}_e = \begin{bmatrix} \sigma_{e,x} & 0 \\ 0 & \sigma_{e,y} \end{bmatrix}$, $\bar{\bar{\sigma}}_m = \begin{bmatrix} \sigma_{m,x} & 0 \\ 0 & \sigma_{m,y} \end{bmatrix}$ and $\bar{\bar{\sigma}}_\chi = \begin{bmatrix} \sigma_{\chi,x} & 0 \\ 0 & \sigma_{\chi,y} \end{bmatrix}$ in this work, where each matrix element can be taken to be arbitrary values. In principle, the derivation here can be readily generalized to the cases with other forms of surface conductivities, such as those with nonzero off-diagonal terms, e.g. in systems with broken time reversal symmetry. Remarkably, we reveal the possibility to achieve the perfect polarization transformation between linearly polarized light in the limit when the chiral surface conductivity dominates. To be specific, this exotic chiral phenomenon can occur either in the transmission or reflection mode, if the chiral surface conductivity fulfills the condition of $\sigma_{\chi,x} \cdot \sigma_{\chi,y} = 1$ or $\sigma_{\chi,x} \cdot \sigma_{\chi,y} = -1$, respectively.

**Results and discussions**

*<u>Reflection and transmission from a chiral metasurface with arbitrary $\bar{\bar{\sigma}}_e$, $\bar{\bar{\sigma}}_m$ and $\bar{\bar{\sigma}}_\chi$</u>*

We begin with the analysis of the scattering coefficients due to a chiral metasurface in the framework of the full Maxwell equations [48-50]. Due to the chiral response of twisted atomic bilayers, the reflected and transmitted light generally has both the transverse magnetic (TM, or *p*-polarized) and electric (TE, or *s*-polarized) components, irrespective of the polarization of incident light; see the schematic illustration in Fig. 1. Without loss of generality, here we set the chiral metasurface to be located at the boundary (at $z = 0$) between region 1 and region 2. For region 1 (region 2), its permittivity and permeability are $\varepsilon_1$ and $\mu_1$ ($\varepsilon_2$ and $\mu_2$), respectively. Below we present the general solution for the reflection and transmission under the incidence of TM waves, while the corresponding solution under the incidence of TE waves is given in the Appendix A.

Under the incidence of TM waves, the magnetic field of incident waves has the following form

$$\bar{H}_i = \hat{y}e^{i\bar{k}_1 \cdot \bar{r}} = \hat{y}e^{ik_x x + ik_{1z} z}, \tag{3}$$

where $\bar{k}_1 = \hat{x}k_x + \hat{z}k_{1z}$ is the wavevector of incident light in region 1. For simplicity, the incident plane is set to be parallel to the $xz$ plane. Since $\nabla \times \bar{H} = -i\omega\varepsilon\bar{E}$ in a homogeneous isotropic media, the corresponding electric field of incident light is

$$\bar{E}_i = -\frac{k_1}{\omega\varepsilon_1}(\hat{z}\frac{k_x}{k_1} - \hat{x}\frac{k_{1z}}{k_1})e^{ik_x x + ik_{1z} z}, \tag{4}$$

The reflected and transmitted fields then take the following forms

$$\bar{H}_r^{TM} = \hat{y}e^{ik_x x - ik_{1z} z} \cdot R_{r,TM}^{i,TM}, \tag{5}$$

$$\bar{E}_r^{TM} = -\frac{k_1}{\omega\varepsilon_1}(\hat{z}\frac{k_x}{k_1} + \hat{x}\frac{k_{1z}}{k_1})e^{ik_x x - ik_{1z} z} \cdot R_{r,TM}^{i,TM}, \tag{6}$$

$$\bar{E}_r^{TE} = \frac{k_1}{\omega\varepsilon_1}\hat{y}e^{ik_x x - ik_{1z} z} \cdot R_{r,TE}^{i,TM}, \tag{7}$$

$$\bar{H}_r^{TE} = (\hat{z}\frac{k_x}{k_1} + \hat{x}\frac{k_{1z}}{k_1})e^{ik_x x - ik_{1z} z} \cdot R_{r,TE}^{i,TM}, \tag{8}$$

$$\bar{H}_t^{TM} = \hat{y}e^{ik_x x + ik_{2z} z} \cdot T_{t,TM}^{i,TM}, \tag{9}$$

$$\bar{E}_t^{TM} = -\frac{k_2}{\omega\varepsilon_2}(\hat{z}\frac{k_x}{k_2} - \hat{x}\frac{k_{2z}}{k_2})e^{ik_x x + ik_{2z} z} \cdot T_{t,TM}^{i,TM}, \tag{10}$$

$$\bar{E}_t^{TE} = \frac{k_2}{\omega\varepsilon_2}\hat{y}e^{ik_x x + ik_{2z} z} \cdot T_{t,TE}^{i,TM}, \tag{11}$$

$$\bar{H}_t^{TE} = (\hat{z}\frac{k_x}{k_2} - \hat{x}\frac{k_{2z}}{k_2})e^{ik_x x + ik_{2z} z} \cdot T_{t,TE}^{i,TM}, \tag{12}$$

In above equations, $\bar{k}_2 = \hat{x}k_x + \hat{z}k_{2z}$ is the wavevector of transmitted light in region 2, and we have four unknown coefficients, namely two reflection coefficients ($R_{r,TM}^{i,TM}$ and $R_{r,TE}^{i,TM}$) and two transmission coefficients ($T_{t,TM}^{i,TM}$, $T_{t,TE}^{i,TM}$). The superscript and subscript of denotes the incident polarization (i, TE or i, TM), and that of the reflected (r) and transmitted (t) light, respectively.

The above scattering coefficients can be solved by enforcing the boundary conditions [51-53], i.e. by substituting equations (3-12) into the boundary conditions governed by equations (1-2). Then we have

$$\begin{bmatrix} A_1^{i,TM} & B_1^{i,TM} & C_1^{i,TM} & D_1^{i,TM} \\ A_2^{i,TM} & B_2^{i,TM} & C_2^{i,TM} & D_2^{i,TM} \\ A_3^{i,TM} & B_3^{i,TM} & C_3^{i,TM} & D_3^{i,TM} \\ A_4^{i,TM} & B_4^{i,TM} & C_4^{i,TM} & D_4^{i,TM} \end{bmatrix} \begin{bmatrix} R_{r,TM}^{i,TM} \\ T_{t,TM}^{i,TM} \\ R_{r,TE}^{t,TM} \\ T_{t,TE}^{i,TM} \end{bmatrix} = \begin{bmatrix} F_1^{i,TM} \\ F_2^{i,TM} \\ F_3^{i,TM} \\ F_4^{i,TM} \end{bmatrix}, \qquad (13)$$

where

$$\begin{bmatrix} A_1^{i,TM} & B_1^{i,TM} & C_1^{i,TM} & D_1^{i,TM} \\ A_2^{i,TM} & B_2^{i,TM} & C_2^{i,TM} & D_2^{i,TM} \\ A_3^{i,TM} & B_3^{i,TM} & C_3^{i,TM} & D_3^{i,TM} \\ A_4^{i,TM} & B_4^{i,TM} & C_4^{i,TM} & D_4^{i,TM} \end{bmatrix} = \begin{bmatrix} \frac{k_{1z}}{\omega\varepsilon_1} + \sigma_{m,y} & \frac{k_{2z}}{\omega\varepsilon_2} + \sigma_{m,y} & -\sigma_{\chi,y}\frac{k_1}{\omega\varepsilon_1} & -\sigma_{\chi,y}\frac{k_2}{\omega\varepsilon_2} \\ \sigma_{\chi,x}\frac{k_{1z}}{\omega\varepsilon_1} & -\sigma_{\chi,x}\frac{k_{2z}}{\omega\varepsilon_2} & \frac{k_1}{\omega\varepsilon_1} + \sigma_{m,x}\frac{k_{1z}}{k_1} & -\frac{k_2}{\omega\varepsilon_2} - \sigma_{m,x}\frac{k_{2z}}{k_2} \\ 1 + \sigma_{e,x}\frac{k_{z1}}{\omega\varepsilon_1} & -1 - \sigma_{e,x}\frac{k_{2z}}{\omega\varepsilon_2} & -\sigma_{\chi,x}\frac{k_{1z}}{k_1} & \sigma_{\chi,x}\frac{k_{2z}}{k_2} \\ \sigma_{\chi,y} & \sigma_{\chi,y} & \frac{k_{1z}}{k_1} + \sigma_{e,y}\frac{k_1}{\omega\varepsilon_1} & \frac{k_{2z}}{k_2} + \sigma_{e,y}\frac{k_2}{\omega\varepsilon_2} \end{bmatrix}, \quad (14)$$

$$\begin{bmatrix} F_1^{i,TM} \\ F_2^{i,TM} \\ F_3^{i,TM} \\ F_4^{i,TM} \end{bmatrix} = \begin{bmatrix} \frac{k_{1z}}{\omega\varepsilon_1} - \sigma_{m,y} \\ \sigma_{\chi,x}\frac{k_{1z}}{\omega\varepsilon_1} \\ -1 + \sigma_{e,x}\frac{k_{1z}}{\omega\varepsilon_1} \\ -\sigma_{\chi,x} \end{bmatrix}. \qquad (15)$$

The analytical expressions for $R_{r,TM}^{i,TM}$, $R_{r,TE}^{i,TM}$, $T_{t,TM}^{i,TM}$ and $T_{t,TE}^{i,TM}$ are obtained by solving equation (13);

$$R_{r,TM}^{i,TM} = \frac{(\alpha_1^{i,TM} Q_1^{i,TM} - \beta_1^{i,TM} N_1^{i,TM})/G^{i,TM}}{S^{i,TM}}, \qquad (16)$$

$$T_{t,TM}^{i,TM} = \frac{(\beta_1^{i,TM} M_1^{i,TM} - \alpha_1^{i,TM} P_1^{i,TM})/G^{i,TM}}{S^{i,TM}}, \qquad (17)$$

$$R_{r,TE}^{i,TM} = \frac{(\alpha_2^{i,TM} Q_2^{i,TM} - \beta_2^{i,TM} N_2^{i,TM})/K^{i,TM}}{S^{i,TM}}, \qquad (18)$$

$$T_{t,TE}^{i,TM} = \frac{(\beta_2^{i,TM} M_2^{i,TM} - \alpha_2^{i,TM} P_2^{i,TM})/K^{i,TM}}{S^{i,TM}}, \qquad (19)$$

where

$$\alpha_1^{i,TM} = (F_1^{i,TM} C_2^{i,TM} - F_2^{i,TM} C_1^{i,TM})(D_3^{i,TM} C_4^{i,TM} - D_4^{i,TM} C_3^{i,TM}) - (F_3^{i,TM} C_4^{i,TM} - F_4^{i,TM} C_3^{i,TM})(D_1^{i,TM} C_2^{i,TM} - D_2^{i,TM} C_1^{i,TM}), \qquad (20)$$

$$\beta_1^{i,TM} = (F_1^{i,TM} D_2^{i,TM} - F_2^{i,TM} D_1^{i,TM})(C_3^{i,TM} D_4^{i,TM} - C_4^{i,TM} D_3^{i,TM}) - (F_3^{i,TM} D_4^{i,TM} - F_4^{i,TM} D_3^{i,TM})(C_1^{i,TM} D_2^{i,TM} - C_2^{i,TM} D_1^{i,TM}), \qquad (21)$$

$$M_1^{i,TM} = (A_1^{i,TM} C_2^{i,TM} - A_2^{i,TM} C_1^{i,TM})(D_3^{i,TM} C_4^{i,TM} - D_4^{i,TM} C_3^{i,TM}) - (A_3^{i,TM} C_4^{i,TM} - A_4^{i,TM} C_3^{i,TM})(D_1^{i,TM} C_2^{i,TM} - D_2^{i,TM} C_1^{i,TM}), \qquad (22)$$

$$N_1^{i,TM} = (B_1^{i,TM} C_2^{i,TM} - B_2^{i,TM} C_1^{i,TM})(D_3^{i,TM} C_4^{i,TM} - D_4^{i,TM} C_3^{i,TM}) - (B_3^{i,TM} C_4^{i,TM} - B_4^{i,TM} C_3^{i,TM})(D_1^{i,TM} C_2^{i,TM} - D_2^{i,TM} C_1^{i,TM}), \qquad (23)$$

$$P_1^{i,TM} = (A_1^{i,TM} D_2^{i,TM} - A_2^{i,TM} D_1^{i,TM})(C_3^{i,TM} D_4^{i,TM} - C_4^{i,TM} D_3^{i,TM}) - (A_3^{i,TM} D_4^{i,TM} - A_4^{i,TM} D_3^{i,TM})(C_1^{i,TM} D_2^{i,TM} - C_2^{i,TM} D_1^{i,TM}), \qquad (24)$$

$$Q_1^{i,TM} = (B_1^{i,TM} D_2^{i,TM} - B_2^{i,TM} D_1^{i,TM})(C_3^{i,TM} D_4^{i,TM} - C_4^{i,TM} D_3^{i,TM}) - (B_3^{i,TM} D_4^{i,TM} - B_4^{i,TM} D_3^{i,TM})(C_1^{i,TM} D_2^{i,TM} - C_2^{i,TM} D_1^{i,TM}), \qquad (25)$$

$$\alpha_2^{i,TM} = (F_1^{i,TM} A_2^{i,TM} - F_2^{i,TM} A_1^{i,TM})(B_3^{i,TM} A_4^{i,TM} - B_4^{i,TM} A_3^{i,TM}) - (F_3^{i,TM} A_4^{i,TM} - F_4^{i,TM} A_3^{i,TM})(B_1^{i,TM} A_2^{i,TM} - B_2^{i,TM} A_1^{i,TM}), \qquad (26)$$

$$\beta_2^{i,TM} = \left(F_1^{i,TM}B_2^{i,TM} - F_2^{i,TM}B_1^{i,TM}\right)\left(A_3^{i,TM}B_4^{i,TM} - A_4^{i,TM}B_3^{i,TM}\right) - \left(F_3^{i,TM}B_4^{i,TM} - F_4^{i,TM}B_3^{i,TM}\right)\left(A_1^{i,TM}B_2^{i,TM} - A_2^{i,TM}B_1^{i,TM}\right), \quad (27)$$

$$M_2^{i,TM} = \left(C_1^{i,TM}A_2^{i,TM} - C_2^{i,TM}A_1^{i,TM}\right)\left(B_3^{i,TM}A_4^{i,TM} - B_4^{i,TM}A_3^{i,TM}\right) - \left(C_3^{i,TM}A_4^{i,TM} - C_4^{i,TM}A_3^{i,TM}\right)\left(B_1^{i,TM}A_2^{i,TM} - B_2^{i,TM}A_1^{i,TM}\right), \quad (28)$$

$$N_2^{i,TM} = \left(D_1^{i,TM}A_2^{i,TM} - D_2^{i,TM}A_1^{i,TM}\right)\left(B_3^{i,TM}A_4^{i,TM} - B_4^{i,TM}A_3^{i,TM}\right) - \left(D_3^{i,TM}A_4^{i,TM} - D_4^{i,TM}A_3^{i,TM}\right)\left(B_1^{i,TM}A_2^{i,TM} - B_2^{i,TM}A_1^{i,TM}\right), \quad (29)$$

$$P_2^{i,TM} = \left(C_1^{i,TM}B_2^{i,TM} - C_2^{i,TM}B_1^{i,TM}\right)\left(A_3^{i,TM}B_4^{i,TM} - A_4^{i,TM}B_3^{i,TM}\right) - \left(C_3^{i,TM}B_4^{i,TM} - C_4^{i,TM}B_3^{i,TM}\right)\left(A_1^{i,TM}B_2^{i,TM} - A_2^{i,TM}B_1^{i,TM}\right), \quad (30)$$

$$Q_2^{i,TM} = \left(D_1^{i,TM}B_2^{i,TM} - D_2^{i,TM}B_1^{i,TM}\right)\left(A_3^{i,TM}B_4^{i,TM} - A_4^{i,TM}B_3^{i,TM}\right) - \left(D_3^{i,TM}B_4^{i,TM} - D_4^{i,TM}B_3^{i,TM}\right)\left(A_1^{i,TM}B_2^{i,TM} - A_2^{i,TM}B_1^{i,TM}\right), \quad (31)$$

$$G^{i,TM} = \left(C_1^{i,TM}D_2^{i,TM} - C_2^{i,TM}D_1^{i,TM}\right)\left(C_3^{i,TM}D_4^{i,TM} - C_4^{i,TM}D_3^{i,TM}\right), \quad (32)$$

$$K^{i,TM} = \left(B_1^{i,TM}A_2^{i,TM} - B_2^{i,TM}A_1^{i,TM}\right)\left(A_3^{i,TM}B_4^{i,TM} - A_4^{i,TM}B_3^{i,TM}\right), \quad (33)$$

$$S^{i,TM} = \frac{M_1^{i,TM}Q_1^{i,TM} - P_1^{i,TM}N_1^{i,TM}}{G^{i,TM}} = \frac{M_2^{i,TM}Q_2^{i,TM} - P_2^{i,TM}N_2^{i,TM}}{K^{i,TM}}. \quad (34)$$

Similarly, for the incidence of TE waves, the reflection coefficients ($R_{r,TE}^{i,TE}$, $R_{r,TM}^{i,TE}$) and transmission coefficients ($T_{t,TE}^{i,TE}$, $T_{t,TM}^{i,TE}$) can be obtained by following a similar procedure; see details in the Appendix A.

### *Perfect polarization transformation for the transmitted light*

Although the general scattering coefficients for the far-field response of chiral metasurfaces are obtained above, these expressions are complicated and hamper the intuitive understanding of electromagnetic phenomena. To elucidate emerging chiral phenomena, several limiting cases are considered below. First, we assume that the chiral metasurface is located in a symmetric environment (e.g. vacuum), that is, $\varepsilon_1 = \varepsilon_2 = \varepsilon$ and $\mu_1 = \mu_2 = \mu$. Hence, we have $k_1 = k_2 = k$. Second, we let $\bar{\bar{\sigma}}_e = 0$ and $\bar{\bar{\sigma}}_m = 0$. In other words, we consider the phenomena when the chiral surface conductivity dominates over other conductivities in the problem. With these simplifications, the complexity for these reflection and transmission coefficients can be significantly reduced, as detailed in the subsequent analysis.

We consider the incidence of TM waves in Fig. 2. By applying the above simplification conditions into equations (16-34), we readily have

$$R_{r,TM}^{i,TM} = \frac{-4\sigma_{\chi,x}\sigma_{\chi,y}\left(\sigma_{\chi,y}+\frac{k_z^2}{k^2}\sigma_{\chi,x}\right)\left(\sigma_{\chi,y}-\frac{k_z^2}{k^2}\sigma_{\chi,x}\right)}{\left(\sigma_{\chi,y}+\frac{k_z^2}{k^2}\sigma_{\chi,x}\right)^2(1+\sigma_{\chi,x}\sigma_{\chi,y})^2-\left(\sigma_{\chi,y}-\frac{k_z^2}{k^2}\sigma_{\chi,x}\right)^2(1-\sigma_{\chi,x}\sigma_{\chi,y})^2},\tag{35}$$

$$T_{t,TM}^{i,TM} = \frac{4\frac{k_z^2}{k^2}\sigma_{\chi,x}\sigma_{\chi,y}(1+\sigma_{\chi,x}\sigma_{\chi,y})(1-\sigma_{\chi,x}\sigma_{\chi,y})}{\left(\sigma_{\chi,y}+\frac{k_z^2}{k^2}\sigma_{\chi,x}\right)^2(1+\sigma_{\chi,x}\sigma_{\chi,y})^2-\left(\sigma_{\chi,y}-\frac{k_z^2}{k^2}\sigma_{\chi,x}\right)^2(1-\sigma_{\chi,x}\sigma_{\chi,y})^2},\tag{36}$$

$$R_{r,TE}^{i,TM} = \frac{-4\frac{k_z}{k}\sigma_{\chi,x}\sigma_{\chi,y}(1-\sigma_{\chi,x}\sigma_{\chi,y})\left(\sigma_{\chi,y}-\frac{k_z^2}{k^2}\sigma_{\chi,x}\right)}{\left(\sigma_{\chi,y}+\frac{k_z^2}{k^2}\sigma_{\chi,x}\right)^2(1+\sigma_{\chi,x}\sigma_{\chi,y})^2-\left(\sigma_{\chi,y}-\frac{k_z^2}{k^2}\sigma_{\chi,x}\right)^2(1-\sigma_{\chi,x}\sigma_{\chi,y})^2},\tag{37}$$

$$T_{t,TE}^{i,TM} = \frac{-4\frac{k_z}{k}\sigma_{\chi,x}\sigma_{\chi,y}(1+\sigma_{\chi,x}\sigma_{\chi,y})\left(\sigma_{\chi,y}+\frac{k_z^2}{k^2}\sigma_{\chi,x}\right)}{\left(\sigma_{\chi,y}+\frac{k_z^2}{k^2}\sigma_{\chi,x}\right)^2(1+\sigma_{\chi,x}\sigma_{\chi,y})^2-\left(\sigma_{\chi,y}-\frac{k_z^2}{k^2}\sigma_{\chi,x}\right)^2(1-\sigma_{\chi,x}\sigma_{\chi,y})^2},\tag{38}$$

Interestingly, from equations (36-37), we always have $T_{t,TM}^{i,TM} = R_{r,TE}^{i,TM} = 0$, if

$$\sigma_{\chi,x}\sigma_{\chi,y} = 1 \tag{39}$$

Under this scenario, since $T_{t,TM}^{i,TM} = 0$ and $T_{t,TE}^{i,TM} \neq 0$, the transmitted light always has the polarization different from the incident light, indicating the occurrence of polarization transformation during the transmission process.

Moreover, when equation (39) is satisfied, we further have $R_{r,TM}^{i,TM} = 0$ and $|T_{t,TE}^{i,TM}| = 1$ from equations (35, 38), if

$$\frac{\sigma_{\chi,y}}{\sigma_{\chi,x}} = \frac{k_z^2}{k^2} \tag{40}$$

Remarkably, $|T_{t,TE}^{i,TM}| = 1$ indicates the occurrence of the *perfect* linear-polarization transformation between the incident and transmitted light under the incidence of TM waves; see the schematic illustration in Fig. 2(a). For clarity, the reflection and transmission coefficients are shown as a function of the incident angle $\theta_i$ in Fig. 2(b), where $\sigma_{\chi,x} = 2$ and $\sigma_{\chi,y} = 0.5$ are chosen to satisfy equation (39). Figure 2(b) shows that $T_{t,TM}^{i,TM} = R_{r,TE}^{i,TM} = 0$, irrespective of the incident angle. Moreover, we have $R_{r,TM}^{i,TM} = 0$ and $|T_{t,TE}^{i,TM}| = 1$ at $\theta_i = 60^o$ in Fig. 2(b). We denote this critical incident angle as $\theta_c$ at which the perfect polarization transformation occurs. From the geometry setup in Fig. 2(a) and equation (40), we have $\cos\theta_c = k_z/k = \sqrt{\frac{\sigma_{\chi,y}}{\sigma_{\chi,x}}}$. As such, the value of $\theta_c$ is simply determined by the ratio between $\sigma_{\chi,x}$ and $\sigma_{\chi,y}$, as shown in Fig. 2(c).

In addition, when $\sigma_{\chi,x}\sigma_{\chi,y} = 1$, we have $|T_{t,TE}^{i,TM}| \to 0$ and $R_{r,TM}^{i,TM} \to -1$ if $\theta_i \to 90^o$. This phenomenon indicates the occurrence of total reflection from the chiral metasurface under the grazing incidence, which almost has no transformation of the polarization.

*__Perfect polarization transformation for the reflected light__*

Upon close inspection of equations (35-38), we find that the perfect polarization transformation can also happen for the reflected light under the incidence of TM waves [Fig. 3(a)], if

$$\sigma_{\chi,x}\sigma_{\chi,y} = -1, \tag{45}$$

$$\frac{\sigma_{\chi,y}}{\sigma_{\chi,x}} = -\frac{k_z^2}{k^2}, \tag{46}$$

To be specific, if $\sigma_{\chi,x}\sigma_{\chi,y} = -1$, we always have $T_{t,TM}^{i,TM} = T_{t,TE}^{i,TM} = 0$ in equations (36, 38) under the incidence of TM waves [Fig. 3(b)]. Under this scenario, the total reflection happens, irrespective of the incident angle.

Furthermore, when equation (45-46) are simultaneously fulfilled, we further have $R_{r,TM}^{i,TM} = 0$ & $|R_{r,TE}^{i,TM}| = 1$ in equations (35, 37) at the critical incident angle $\cos\theta_c = \sqrt{-\frac{\sigma_{\chi,y}}{\sigma_{\chi,x}}}$. In other words, the perfect polarization transformation between the incident and reflected light occurs at this critical incident angle. For example, if $\sigma_{\chi,x} = 2$ and $\sigma_{\chi,y} = -0.5$, we have $|R_{r,TE}^{i,TM}| = 1$ in Fig. 3(b) at $\theta_c = 60º$. In addition, similar grazing angle phenomenon is also observed in Fig. 3(b), as that discussed for the transmission mode in Fig. 2(b).

The critical incident angle is also shown as a function of $\frac{\sigma_{\chi,y}}{\sigma_{\chi,x}}$ in Fig. 3(c). No matter for the reflected or transmitted light, both phenomena of the perfect polarization transformation in Figs. 2-3 occur at the critical angle of $\cos\theta_c = \sqrt{|\frac{\sigma_{\chi,y}}{\sigma_{\chi,x}}|}$. As such, all critical incident angles in Figs. 2(c)&3(c) decrease from 90º to 0º when $|\frac{\sigma_{\chi,y}}{\sigma_{\chi,x}}|$ increases from 0 to 1.

Last but not least, we find that $R_{r,TE}^{i,TM} = -R_{r,TM}^{i,TE}$, $T_{t,TE}^{i,TM} = -T_{t,TM}^{i,TE}$, $R_{r,TM}^{i,TM} = +R_{r,TE}^{i,TE}$, and $T_{t,TM}^{i,TM} = +T_{t,TE}^{i,TE}$, if the above simplification conditions are adopted. Therefore, the exotic chiral phenomena in Figs. 2-3 can also occur under the incidence of TE waves, and the polarization of the incident light actually has no influence on the revealed phenomenon of perfect linear-polarization transformation.

**Conclusion**

In conclusion, we have systematically investigated the chiral optics from metasurfaces with a chiral surface conductivity under the incidence of linearly polarized light. We have found an exotic chiral phenomenon, namely the perfect polarization transformation between linearly polarized light from these chiral metasurfaces. To be specific, if $\sigma_{\chi,x}\sigma_{\chi,y} = 1$, such a chiral metasurface can facilitate the realization of total transmittance at a critical incident angle, where the polarization of all transmitted light is distinct from the incident light. In contrast, if $\sigma_{\chi,x}\sigma_{\chi,y} = -1$, this type of chiral metasurfaces can be exploited to achieve total reflection instead, irrespective of the incident angle; moreover, the polarization of all reflected light can be made different from the incident light at the critical incident angle. Our work indicates the rich fundamental physics in twisted atomic bilayers and chiral metasurfaces, which warrants further exploration and could find diverse chiral optics applications.

**ACKNOWLEDGMENTS**
The work was sponsored by the National Natural Science Foundation of China (NNSFC) under Grants No. 61625502, No.11961141010, and No. 61975176, the Top-Notch Young Talents Program of China, the Fundamental Research Funds for the Central Universities. TL acknowledges support by the National Science Foundation, NSF/EFRI Grant No. EFRI-1741660.

**APPENDIX A**
***Reflection and transmission from chiral metasurfaces under the incidence of TE waves***

By following a similar procedure in equations (3-34), we can calculate the reflection and transmission of light from a chiral metasurface under the incidence of TE waves. To be specific, the electric field of incident TE waves has

$$\bar{E}_i = \hat{y}e^{i\bar{k}_1 \cdot \bar{r}} = \hat{y}e^{ik_x x + ik_{1z}z}, \tag{A1}$$

where the wavevector of incident light is $\bar{k}_1 = \hat{x}k_x + \hat{z}k_{1z}$. Correspondingly, the magnetic field of incident light has

$$\bar{H}_i = \frac{k_1}{\omega\mu_1}(\hat{z}\frac{k_x}{k_1} - \hat{x}\frac{k_{1z}}{k_1})e^{ik_x x + ik_{1z}z}, \tag{A2}$$

The reflected and transmitted fields then take the following forms

$$\bar{E}_r^{TE} = \hat{y}e^{ik_x x - ik_{1z}z} \cdot R_{r,TE}^{i,TE}, \tag{A3}$$

$$\bar{H}_r^{TE} = \frac{k_1}{\omega\mu_1}(\hat{z}\frac{k_x}{k_1} + \hat{x}\frac{k_{1z}}{k_1})e^{ik_x x - ik_{1z}z} \cdot R_{r,TE}^{i,TE}, \tag{A4}$$

$$\overline{H}_r^{TM} = \frac{k_1}{\omega\mu_1}\hat{y}e^{ik_x x - ik_{1z}z} \cdot R_{r,TM}^{i,TE}, \tag{A5}$$

$$\overline{E}_r^{TM} = -(\hat{z}\frac{k_x}{k_1} + \hat{x}\frac{k_{1z}}{k_1})e^{ik_x x - ik_{1z}z} \cdot R_{r,TM}^{i,TE}, \tag{A6}$$

$$\overline{E}_t^{TE} = \hat{y}e^{ik_x x + ik_{2z}z} \cdot T_{t,TE}^{i,TE}, \tag{A7}$$

$$\overline{H}_t^{TE} = \frac{k_2}{\omega\mu_2}(\hat{z}\frac{k_x}{k_2} - \hat{x}\frac{k_{2z}}{k_2})e^{ik_x x + ik_{2z}z} \cdot T_{t,TE}^{i,TE}, \tag{A8}$$

$$\overline{H}_t^{TM} = \frac{k_2}{\omega\mu_2}\hat{y}e^{ik_x x + ik_{2z}z} \cdot T_{t,TM}^{i,TE}, \tag{A9}$$

$$\overline{E}_t^{TM} = -(\hat{z}\frac{k_x}{k_2} - \hat{x}\frac{k_{2z}}{k_2})e^{ik_x x + ik_{2z}z} \cdot T_{t,TM}^{i,TE}. \tag{A10}$$

where the wavevector of transmitted light is $\bar{k}_2 = \hat{x}k_x + \hat{z}k_{2z}$. The four coefficients, namely $R_{r,TE}^{i,TE}$ and $R_{r,TM}^{i,TE}$ (reflection coefficients) and $T_{t,TE}^{i,TE}$ and $T_{t,TM}^{i,TE}$ (transmission coefficients), can be solved by enforcing the boundary conditions of equations (1-2) in the main text. By substituting equations (A1-A10) into equations (1-2), we obtain the relation of these coefficients as follows

$$\begin{bmatrix} A_1^{i,TM} & B_1^{i,TM} & C_1^{i,TM} & D_1^{i,TM} \\ A_2^{i,TM} & B_2^{i,TM} & C_2^{i,TM} & D_2^{i,TM} \\ A_3^{i,TM} & B_3^{i,TM} & C_3^{i,TM} & D_3^{i,TM} \\ A_4^{i,TM} & B_4^{i,TM} & C_4^{i,TM} & D_4^{i,TM} \end{bmatrix} \begin{bmatrix} R_{r,TM}^{i,TM} \\ T_{t,TM}^{i,TM} \\ R_{r,TE}^{i,TM} \\ T_{t,TE}^{i,TM} \end{bmatrix} = \begin{bmatrix} F_1^{i,TM} \\ F_2^{i,TM} \\ F_3^{i,TM} \\ F_4^{i,TM} \end{bmatrix}, \tag{A11}$$

where

$$\begin{bmatrix} A_1^{i,TE} & B_1^{i,TE} & C_1^{i,TE} & D_1^{i,TE} \\ A_2^{i,TE} & B_2^{i,TE} & C_2^{i,TE} & D_2^{i,TE} \\ A_3^{i,TE} & B_3^{i,TE} & C_3^{i,TE} & D_3^{i,TE} \\ A_4^{i,TE} & B_4^{i,TE} & C_4^{i,TE} & D_4^{i,TE} \end{bmatrix} = \begin{bmatrix} 1 + \sigma_{m,x}\frac{k_{1z}}{\omega\mu_1} & -1 - \sigma_{m,x}\frac{k_{2z}}{\omega\mu_2} & \sigma_{\chi,x}\frac{k_{1z}}{k_1} & -\sigma_{\chi,x}\frac{k_{2z}}{k_2} \\ \sigma_{\chi,y} & \sigma_{\chi,x} & -\sigma_{m,y}\frac{k_1}{\omega\mu_1} - \frac{k_{1z}}{k_1} & -\sigma_{m,y}\frac{k_2}{\omega\mu_2} - \frac{k_{2z}}{k_2} \\ \sigma_{e,y} + \frac{k_{1z}}{\omega\mu_1} & \sigma_{e,y} + \frac{k_{2z}}{\omega\mu_2} & \sigma_{\chi,y}\frac{k_1}{\omega\mu_1} & \sigma_{\chi,y}\frac{k_2}{\omega\mu_2} \\ \sigma_{\chi,x}\frac{k_{1z}}{\omega\mu_1} & -\sigma_{\chi,x}\frac{k_{2z}}{\omega\mu_2} & -\frac{k_1}{\omega\mu_1} - \sigma_{e,x}\frac{k_{1z}}{k_1} & \frac{k_2}{\omega\mu_2} + \sigma_{e,x}\frac{k_{2z}}{k_2} \end{bmatrix}, \tag{A12}$$

$$\begin{bmatrix} F_1^{i,TE} \\ F_2^{i,TE} \\ F_3^{i,TE} \\ F_4^{i,TE} \end{bmatrix} = \begin{bmatrix} -1 + \sigma_{m,x}\frac{k_{1z}}{\omega\mu_1} \\ -\sigma_{\chi,x} \\ \frac{k_{1z}}{\omega\mu_1} - \sigma_{e,y} \\ \sigma_{\chi,x}\frac{k_{1z}}{\omega\mu_1} \end{bmatrix}. \tag{A13}$$

$R_{r,TE}^{i,TE}$, $R_{r,TM}^{i,TE}$, $T_{t,TE}^{i,TE}$ and $T_{t,TM}^{i,TE}$ then can be solved from equation (A11). They are expressed as follows:

$$R_{r,TE}^{i,TE} = \frac{(\alpha_1^{i,TE}Q_1^{i,TE} - \beta_1^{i,TE}N_1^{i,TE})/G^{i,TE}}{S^{i,TE}}, \tag{A14}$$

$$T_{t,TE}^{i,TE} = \frac{(\beta_1^{i,TE}M_1^{i,TE} - \alpha_1^{i,TE}P_1^{i,TE})/G^{i,TE}}{S^{i,TE}}, \tag{A15}$$

$$R_{r,TM}^{i,TE} = \frac{(\alpha_2^{i,TE}Q_2^{i,TE} - \beta_2^{i,TE}N_2^{i,TE})/K^{i,TE}}{S^{i,TE}}, \tag{A16}$$

$$T_{t,TM}^{i,TE} = \frac{(\beta_2^{i,TE}M_2^{i,TE} - \alpha_2^{i,TE}P_2^{i,TE})/K^{i,TE}}{S^{i,TE}}, \tag{A17}$$

where,

$$\alpha_1^{i,TE} = (F_1^{i,TE}C_2^{i,TE} - F_2^{i,TE}C_1^{i,TE})(D_3^{i,TE}C_4^{i,TE} - D_4^{i,TE}C_3^{i,TE}) - (F_3^{i,TE}C_4^{i,TE} - F_4^{i,TE}C_3^{i,TE})(D_1^{i,TE}C_2^{i,TE} - D_2^{i,TE}C_1^{i,TE}), \tag{A18}$$

$$\beta_1^{i,TE} = (F_1^{i,TE}D_2^{i,TE} - F_2^{i,TE}D_1^{i,TE})(C_3^{i,TE}D_4^{i,TE} - C_4^{i,TE}D_3^{i,TE}) - (F_3^{i,TE}D_4^{i,TE} - F_4^{i,TE}D_3^{i,TE})(C_1^{i,TE}D_2^{i,TE} - C_2^{i,TE}D_1^{i,TE}), \tag{A19}$$

$$M_1^{i,TE} = (A_1^{i,TE}C_2^{i,TE} - A_2^{i,TE}C_1^{i,TE})(D_3^{i,TE}C_4^{i,TE} - D_4^{i,TE}C_3^{i,TE}) - (A_3^{i,TE}C_4^{i,TE} - A_4^{i,TE}C_3^{i,TE})(D_1^{i,TE}C_2^{i,TE} - D_2^{i,TE}C_1^{i,TE}), \tag{A20}$$

$$N_1^{i,TE} = (B_1^{i,TE}C_2^{i,TE} - B_2^{i,TE}C_1^{i,TE})(D_3^{i,TE}C_4^{i,TE} - D_4^{i,TE}C_3^{i,TE}) - (B_3^{i,TE}C_4^{i,TE} - B_4^{i,TE}C_3^{i,TE})(D_1^{i,TE}C_2^{i,TE} - D_2^{i,TE}C_1^{i,TE}), \tag{A21}$$

$$P_1^{i,TE} = (A_1^{i,TE}D_2^{i,TE} - A_2^{i,TE}D_1^{i,TE})(C_3^{i,TE}D_4^{i,TE} - C_4^{i,TE}D_3^{i,TE}) - (A_3^{i,TE}D_4^{i,TE} - A_4^{i,TE}D_3^{i,TE})(C_1^{i,TE}D_2^{i,TE} - C_2^{i,TE}D_1^{i,TE}), \tag{A22}$$

$$Q_1^{i,TE} = (B_1^{i,TE}D_2^{i,TE} - B_2^{i,TE}D_1^{i,TE})(C_3^{i,TE}D_4^{i,TE} - C_4^{i,TE}D_3^{i,TE}) - (B_3^{i,TE}D_4^{i,TE} - B_4^{i,TE}D_3^{i,TE})(C_1^{i,TE}D_2^{i,TE} - C_2^{i,TE}D_1^{i,TE}), \tag{A23}$$

$$\alpha_2^{i,TE} = (F_1^{i,TE}A_2^{i,TE} - F_2^{i,TE}A_1^{i,TE})(B_3^{i,TE}A_4^{i,TE} - B_4^{i,TE}A_3^{i,TE}) - (F_3^{i,TE}A_4^{i,TE} - F_4^{i,TE}A_3^{i,TE})(B_1^{i,TE}A_2^{i,TE} - B_2^{i,TE}A_1^{i,TE}), \tag{A24}$$

$$\beta_2^{i,TE} = (F_1^{i,TE}B_2^{i,TE} - F_2^{i,TE}B_1^{i,TE})(A_3^{i,TE}B_4^{i,TE} - A_4^{i,TE}B_3^{i,TE}) - (F_3^{i,TE}B_4^{i,TE} - F_4^{i,TE}B_3^{i,TE})(A_1^{i,TE}B_2^{i,TE} - A_2^{i,TE}B_1^{i,TE}), \tag{A25}$$

$$M_2^{i,TE} = (C_1^{i,TE}A_2^{i,TE} - C_2^{i,TE}A_1^{i,TE})(B_3^{i,TE}A_4^{i,TE} - B_4^{i,TE}A_3^{i,TE}) - (C_3^{i,TE}A_4^{i,TE} - C_4^{i,TE}A_3^{i,TE})(B_1^{i,TE}A_2^{i,TE} - B_2^{i,TE}A_1^{i,TE}), \tag{A26}$$

$$N_2^{i,TE} = (D_1^{i,TE}A_2^{i,TE} - D_2^{i,TE}A_1^{i,TE})(B_3^{i,TE}A_4^{i,TE} - B_4^{i,TE}A_3^{i,TE}) - (D_3^{i,TE}A_4^{i,TE} - D_4^{i,TE}A_3^{i,TE})(B_1^{i,TE}A_2^{i,TE} - B_2^{i,TE}A_1^{i,TE}), \tag{A27}$$

$$P_2^{i,TE} = (C_1^{i,TE}B_2^{i,TE} - C_2^{i,TE}B_1^{i,TE})(A_3^{i,TE}B_4^{i,TE} - A_4^{i,TE}B_3^{i,TE}) - (C_3^{i,TE}B_4^{i,TE} - C_4^{i,TE}B_3^{i,TE})(A_1^{i,TE}B_2^{i,TE} - A_2^{i,TE}B_1^{i,TE}), \tag{A28}$$

$$Q_2^{i,TE} = (D_1^{i,TE}B_2^{i,TE} - D_2^{i,TE}B_1^{i,TE})(A_3^{i,TE}B_4^{i,TE} - A_4^{i,TE}B_3^{i,TE}) - (D_3^{i,TE}B_4^{i,TE} - D_4^{i,TE}B_3^{i,TE})(A_1^{i,TE}B_2^{i,TE} - A_2^{i,TE}B_1^{i,TE}), \tag{A29}$$

$$G^{i,TE} = (C_1^{i,TE}D_2^{i,TE} - C_2^{i,TE}D_1^{i,TE})(C_3^{i,TE}D_4^{i,TE} - C_4^{i,TE}D_3^{i,TE}), \tag{A30}$$

$$K^{i,TE} = (B_1^{i,TE}A_2^{i,TE} - B_2^{i,TE}A_1^{i,TE})(A_3^{i,TE}B_4^{i,TE} - A_4^{i,TE}B_3^{i,TE}), \tag{A31}$$

$$S^{i,TE} = \frac{M_1^{i,TE}Q_1^{i,TE} - P_1^{i,TE}N_1^{i,TE}}{G^{i,TE}} = \frac{M_2^{i,TE}Q_2^{i,TE} - P_2^{i,TE}N_2^{i,TE}}{K^{i,TE}}, \tag{A32}$$

If the simplification conditions (i.e. $\varepsilon_1 = \varepsilon_2 = \varepsilon$, $\mu_1 = \mu_2 = \mu$, $\bar{\bar{\sigma}}_e = 0$, and $\bar{\bar{\sigma}}_m = 0$) are adopted in the calculation, the expression for the scattering coefficients in equations (A14-A17) can be significantly simplified. Under these simplification conditions, we have

$$R_{r,TE}^{i,TE} = \frac{-4\sigma_{\chi,x}\sigma_{\chi,y}\left(\sigma_{\chi,y} + \frac{k_z^2}{k^2}\sigma_{\chi,x}\right)\left(\sigma_{\chi,y} - \frac{k_z^2}{k^2}\sigma_{\chi,x}\right)}{\left(\sigma_{\chi,y} + \frac{k_z^2}{k^2}\sigma_{\chi,x}\right)^2(1+\sigma_{\chi,x}\sigma_{\chi,y})^2 - \left(\sigma_{\chi,y} - \frac{k_z^2}{k^2}\sigma_{\chi,x}\right)^2(1-\sigma_{\chi,x}\sigma_{\chi,y})^2}, \tag{A33}$$

$$T_{t,TE}^{i,TE} = \frac{4\sigma_{\chi,x}\sigma_{\chi,y}\frac{k_z^2}{k^2}(1+\sigma_{\chi,x}\sigma_{\chi,y})(1-\sigma_{\chi,x}\sigma_{\chi,y})}{\left(\sigma_{\chi,y} + \frac{k_z^2}{k^2}\sigma_{\chi,x}\right)^2(1+\sigma_{\chi,x}\sigma_{\chi,y})^2 - \left(\sigma_{\chi,y} - \frac{k_z^2}{k^2}\sigma_{\chi,x}\right)^2(1-\sigma_{\chi,x}\sigma_{\chi,y})^2}, \tag{A34}$$

$$R_{r,TM}^{i,TE} = \frac{4\frac{k_z}{k}\sigma_{\chi,x}\sigma_{\chi,y}\left(\sigma_{\chi,y} - \frac{k_z^2}{k^2}\sigma_{\chi,x}\right)(1-\sigma_{\chi,x}\sigma_{\chi,y})}{\left(\sigma_{\chi,y} + \frac{k_z^2}{k^2}\sigma_{\chi,x}\right)^2(1+\sigma_{\chi,x}\sigma_{\chi,y})^2 - \left(\sigma_{\chi,y} - \frac{k_z^2}{k^2}\sigma_{\chi,x}\right)^2(1-\sigma_{\chi,x}\sigma_{\chi,y})^2}, \tag{A35}$$

$$T_{t,TM}^{i,TE} = \frac{4\frac{k_z}{k}\sigma_{\chi,x}\sigma_{\chi,y}\left(\sigma_{\chi,y} + \frac{k_z^2}{k^2}\sigma_{\chi,x}\right)(1+\sigma_{\chi,x}\sigma_{\chi,y})}{\left(\sigma_{\chi,y} + \frac{k_z^2}{k^2}\sigma_{\chi,x}\right)^2(1+\sigma_{\chi,x}\sigma_{\chi,y})^2 - \left(\sigma_{\chi,y} - \frac{k_z^2}{k^2}\sigma_{\chi,x}\right)^2(1-\sigma_{\chi,x}\sigma_{\chi,y})^2}. \tag{A36}$$

*These authors contributed equally to this work.
†Corresponding author.
*tlow@umn.edu*
‡Corresponding author.
*hansomchen@zju.edu.cn*
§Corresponding author.
*xiaolinzju@zju.edu.cn*

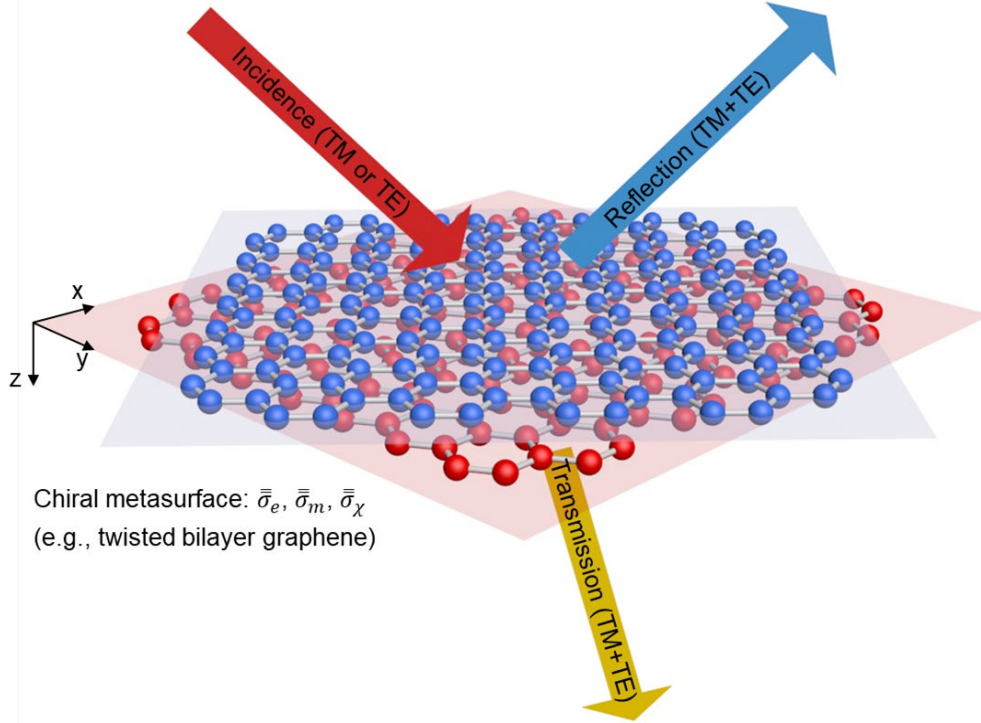

FIG. 1. Schematic of the reflection and transmission from a chiral metasurface under the incidence of linearly polarized waves. The chiral metasurface is featured with an electric surface conductivity $\bar{\bar{\sigma}}_e$, a magnetic surface conductivity $\bar{\bar{\sigma}}_m$, and a chiral surface conductivity $\bar{\bar{\sigma}}_\chi$. The chiral metasurface is located at the boundary between region 1 ($z < 0$) and region 2 ($z > 0$). Due to the chiral response of metasurfaces, the reflected and transmitted waves generally contain both the transverse-electric (TE, or *s*-polarized) and transverse-magnetic (TM, or *p*-polarized) field components.

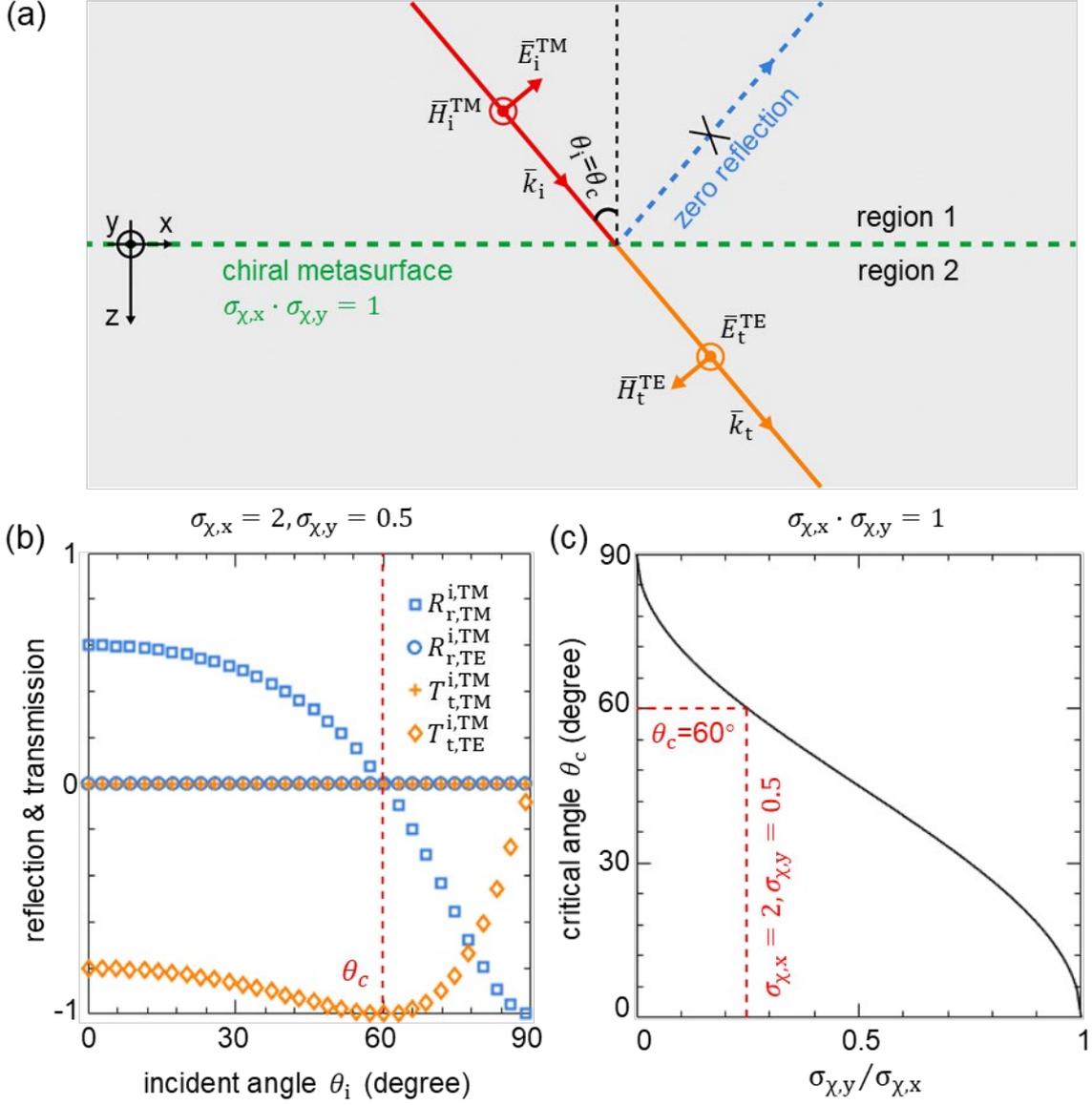

FIG. 2. Perfect polarization transformation between TM (incident) and TE (transmitted) waves. Here the chiral metasurface has $\sigma_{\chi,x}\sigma_{\chi,y}=1$. Regions 1 and 2 are the same dielectric (e.g. vacuum used here). (a) Schematic illustration. (b) Reflection and transmission coefficients as a function of the incident angle $\theta_i$. For conceptual demonstration, $\sigma_{\chi,x}=2$ and $\sigma_{\chi,y}=0.5$ are chosen. If $\sigma_{\chi,x}\sigma_{\chi,y}=1$, $R_{r,TE}^{i,TM}$ and $T_{t,TM}^{i,TM}$ are equal to zero, irrespective of the incident angle. The critical incident angle is denoted as $\theta_c$, at which $R_{r,TM}^{i,TM}=0$ and $|T_{t,TE}^{i,TM}|=1$. (c) $\theta_c$ as a function of $\sigma_{\chi,y}/\sigma_{\chi,x}$.

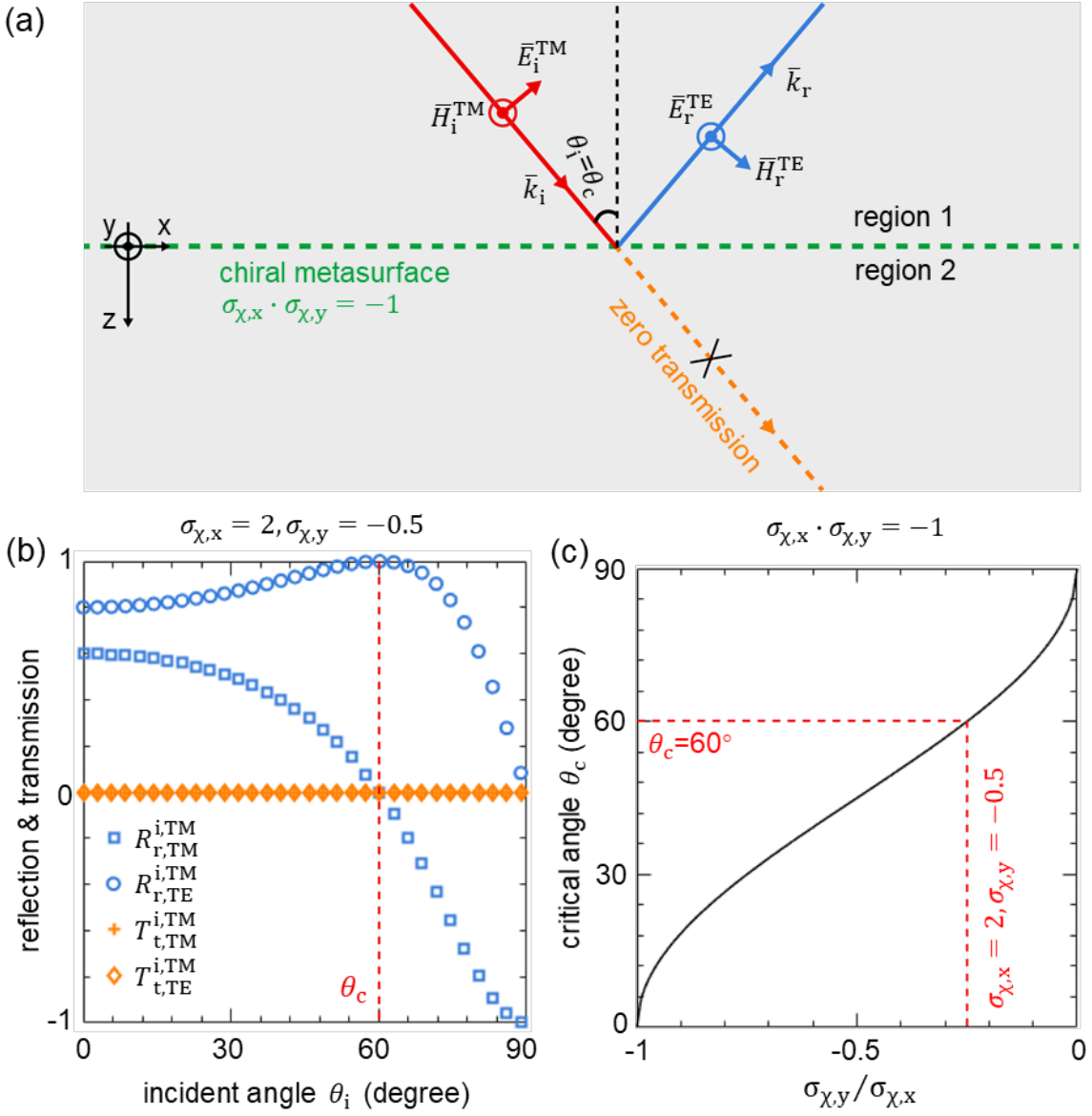

FIG. 3. Perfect polarization transformation between TM (incident) and TE (reflected) waves. The basic structural setup is the same as Fig. 2, except for that the chiral metasurface has $\sigma_{\chi,x}\sigma_{\chi,y} = -1$. (a) Schematic illustration. (b) Reflection and transmission coefficients as a function of $\theta_i$. Here $\sigma_{\chi,x} = 2$ and $\sigma_{\chi,y} = -0.5$ are chosen for conceptual illustration. If $\sigma_{\chi,x}\sigma_{\chi,y} = -1$, $T_{t,TE}^{i,TM}$ and $T_{t,TM}^{i,TM}$ are equal to zero for arbitrary incident angle. The critical incident angle is denoted as $\theta_c$, at which $R_{r,TM}^{i,TM} = 0$ and $|R_{r,TE}^{i,TM}| = 1$. (c) $\theta_c$ as a function of $\sigma_{\chi,y}/\sigma_{\chi,x}$.

17